\documentclass[aps,reprint,article,prx]{revtex4-1}
\usepackage{amsmath}
\usepackage{dcolumn}
\usepackage{graphicx}
\usepackage{float}

\begin{document}
\title{Observation and characterization of the zero energy conductance peak in the vortex core state of FeTe$_{0.55}$Se$_{0.45}$}
\author{Xiaoyu Chen, Mingyang Chen, Wen Duan, Xiyu Zhu, Huan Yang$^*$ and Hai-Hu Wen$^\dag$}

\affiliation{National Laboratory of Solid State Microstructures and Department of Physics, Collaborative Innovation Center of Advanced Microstructures, Nanjing University, Nanjing 210093, China}

\begin{abstract}
The search  for the Majorana fermions in condensed matter physics has attracted much attention, partially because they may be used for the fault-tolerant quantum computation. It has been predicted that the Majorana zero mode may exist in the vortex core of topological superconductors. Recently, many iron-based superconductors are claimed to exhibit a topologically nontrivial surface state, including Fe(Te,Se). Some previous experiments through scanning tunneling microscopy (STM) have found zero-bias conductance peaks (ZBCP) within the vortex cores of Fe(Te,Se). However, our early experimental results have revealed the Caroli-de Gennes-Matricon (CdGM) discrete quantum levels in about 20\% vortices. In many other vortices, we observed a dominant peak locating near zero bias. Here we show further study on the vortex core state of many more vortices in FeTe$_{0.55}$Se$_{0.45}$. In some vortices, if we take a certain criterion of bias voltage window near zero energy, we indeed see a zero energy mode. Some vortices exhibit zero energy bound state peaks with relatively symmetric background, which cannot be interpreted as the CdGM discrete states. The probability for vortices showing the ZBCP lowers down with the increase of magnetic field. Meanwhile it seems that the presence and absence of the ZBCP has no clear relationship with the Te/Se ratio on the surface. Temperature dependence of the spectra reveals that the ZBCP becomes weakened with increasing temperature and disappears at about 4 K. Our results provide a confirmed supplementary to the early claimed zero energy modes within the vortex cores of Fe(Te,Se). Detailed characterization of these zero energy modes versus magnetic field, temperature and spatial distribution of Te/Se will help to clarify its origin.
\end{abstract}

\maketitle
\section{Introduction}
In a type-II superconductor at an external magnetic field stronger than the lower critical field, the quantized vortices will appear with the supercurrent circling around the vortex core. The vortex core region shows a suppression of superconducting order parameter and can be described as a normal-state cylinder with the radius comparable to the coherence length $\xi$. The quasiparticle excitations then take place in the vortex core and form the vortex bound states. The electronic properties of vortex core state were first studied by Caroli, de Gennes and Matricon (CdGM) \cite{CdGM} in 1964, and their calculation shows that the quasiparticles in the vortex core region will form discrete bound states (CdGM states) in quantum limit with the energy level $E_\mu=\pm\mu\Delta^ 2/E_F$ with $\Delta$ the superconducting gap and $E_F$ the Fermi energy. Further theoretical studies by solving the Bogoliubov-de Gennes equations \cite{vortex theory 2} reached the conclusion that $\mu$ = 1/2, 3/2, 5/2 $\cdots$. In many conventional superconductors, the Fermi energy is usually quite large (about several hundred meV to several eV) compared with the superconducting energy gap. Therefore the energy level spacing of the CdGM state is relatively small. In this case, one can see a dominant peak at zero bias in the tunneling spectrum measured at the vortex core center, and the central peak consists of accumulated states near zero energy with low angular momenta $\mu$. When moving away from the vortex core center, the peak splits into two because the bound state peaks at higher eigenstate energies with larger $\mu$ will dominate the contribution to the density of states \cite{vortex theory 2,Hess 1,Hess 2,vortex theory 1}.

FeTe$_{1-x}$Se$_{x}$ is one of the typical iron-based superconductors with maximum critical temperature $T_c$ of about 14.6 K \cite{FTS Tc}. The angle resolved photoemission spectroscopy (ARPES) measurements show that the compound has very small Fermi energy (several meV) \cite{ARPES FTS1,ARPES FTS2}. In our previous study in the vortex cores in FeTe$_{0.55}$Se$_{0.45}$, we can observe clear CdGM states with discrete quantum levels in about 20\% of measured vortices \cite{CdGM NC}. This is due to very small $E_F$ in the material, and the quantum limit condition can be easily satisfied. In some other vortices, we observe a dominant in-gap peak near zero bias on the tunneling spectra.

Recent studies show that FeTe$_{1-x}$Se$_{x}$ may have a topological surface state. Theoretically it was proposed that the downward shifted $p_z$ band due to the Te substitution will cross the $d_{xz}$ band near Fermi energy \cite{ARPES1,FST Band,HuJP}. Because of the strong spin-orbital coupling effect and the parity feature of these bands, a topological surface state exists with a spin-helical texture. This spin-helical feature has been seen by the ARPES measurements which show a Dirac cone like dispersion in the surface band structure \cite{ARPES2}. In addition, a robust zero-bias differential conductance peak is observed at the interstitial excess iron atom on the cleaved surface, and the peak does not split under magnetic fields \cite{SH Pan}. Theoretical explanation about this feature is a Majorana zero mode existing in a quantum anomalous vortex induced by a magnetic impurity when the topological surface state exists in this material \cite{ZQ Wang}. Therefore, FeTe$_{1-x}$Se$_{x}$ may host a topological nontrivial surface state and the surface layer behaves as a topological superconductor due to the intrinsic proximity effect.

The vortex core state of topological superconductors shows different behaviors comparing with the CdGM states in a conventional superconductor. It is predicted that one can find Majorana zero modes in the vortices of two dimensional $p+ip$ superconductors \cite{Read and Green}. By using the proximity effect between an s-wave superconductor and a topological insulator, a 2D topological superconducting state which resembles a $p+ip$ superconductor can be achieved. In this case Majorana zero modes can also exist in the vortex cores \cite{Fu and Kane}. Through scanning tunneling microscopy/spectroscopy (STM/STS) measurement, a zero-bias conductance peak (ZBCP) can be found in the tunneling spectra taken at the vortex cores of topological superconductors. The zero energy mode in the vortex core states of topological superconductors shows different behavior comparing with the CdGM states.

In recent STM measurement of vortices on FeTe$_{0.55}$Se$_{0.45}$, the ZBCPs were observed and these ZBCPs show a non-splitting behavior in space \cite{HJ Gao}. It was also suggested that the CdGM state and the Majorana bound state may differ by a half integer energy level shift in the spectra of the vortex bound states \cite{half intefer level shift}. Later on, a quantized conductance plateau at $2e^2/h$  was also observed \cite{2e2/h FTS}. Theoretically it is predicted that the tunneling into an isolated Majorana mode will give rise to a zero-bias conductance peak with the upper limit height of $2e^2/h$ \cite{quantized plateau 1,quantized plateau 2,quantized plateau 3}. The result suggests the possible existence of the Majorana zero modes in the vortex cores of FeTe$_{0.55}$Se$_{0.45}$. In some other iron-based superconductors, the coexistence of superconductivity and nontrivial topological states were also suggested \cite{P Zhang NP}. Furthermore, the Majorana zero mode and the quantized conductance plateau were also detected in (Li$_{1-x}$Fe$_{x}$)OHFeSe \cite{LiFeOHFeSe 1,LiFeOHFeSe 2}.

In this paper, we report our further study on the vortex core states in FeTe$_{0.55}$Se$_{0.45}$. By measuring hundreds of vortices in the FeTe$_{0.55}$Se$_{0.45}$ single crystals, we also find zero-bias differential conductance peaks on the tunneling spectra measured on some vortices. Furthermore, the influences of the magnetic field, temperature, and the Te concentration on the ZBCPs have also been studied. Our results give useful information for the understanding of the ZBCPs (perhaps due to the Majorana modes) appearing on some of the vortices in FeTe$_{0.55}$Se$_{0.45}$.

\section{Methods}

The single crystals used in this work were grown with the nominal composition of FeTe$_{0.55}$Se$_{0.45}$\cite{Growth}. Fe (99.9\%), Te (99.999\%) and Se (99.999\%) powders were mixed with the ratio of 1.04:0.55:0.45 and then sealed in an evacuated quartz ampoule. The mixture was heated to 1080 $^\circ$C and stayed for 15 hours, then it was slowly cooled down to 400 $^\circ$C at a cooling rate of 3 $^\circ$C/h to obtain single crystals. After that, we annealed the sample at 400 $^{\circ}$C for 20 h in appropriate O$_2$ atmosphere to remove the excess Fe atoms. Then it is quenched in liquid nitrogen. The STM/STS measurements were carried out in the STM (USM-1300, Unisoku Co., Ltd.) system. The FeTe$_{0.55}$Se$_{0.45}$ samples were cleaved in an ultrahigh vacuum with the base pressure of about $1\times10^{-10}$ Torr. Electrochemical etched tungsten tips were used for the STM measurements after electron-beam heating in the ultrahigh vacuum chamber. A typical lock-in technique was used for the STS measurements with an ac modulation of 0.3 mV and the frequency of 831.773 Hz. The voltage offsets of the tunneling spectra were carefully calibrated \cite{CdGM NC}. All data were measured at the temperature about 400 mK except for some cases with temperatures specially stated.

\section{Results}

\begin{figure}[H]
\centering
\includegraphics[width=8cm]{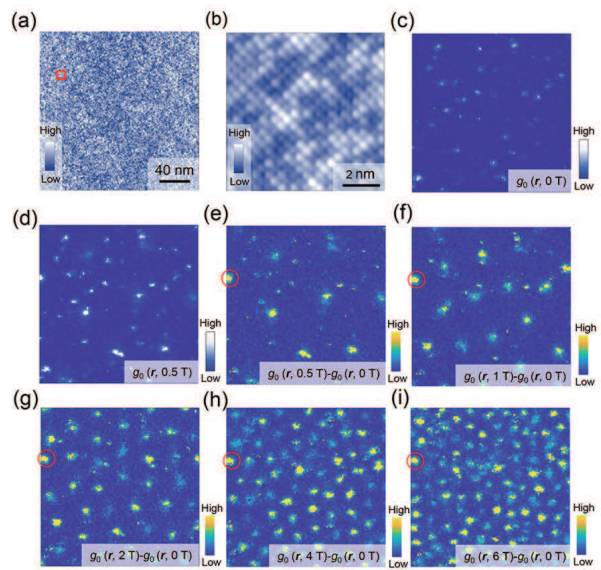}
\caption{Topographic image of the sample surface and differential conductance mappings of vortices under magnetic fields. (a) Topographic image measured on the cleaved surface of FeTe$_{0.55}$Se$_{0.45}$. The setpoint conditions are bias voltage $V_{bias} = 1$ V and tunneling current $I_{set} = 10$ pA. (b) Atomically resolved topography re-measured in the marked square area in (a) with $V_{bias} = 10$ mV and $I_{set} = 100$ pA. (c,d) Zero-bias differential conductance mapping measured in the same area of (a) and at 0 and 0.5 T, respectively. (e-i) Images of the vortices by subtracting the zero-bias differential conductance mapping measured at 0 T [$g_0(\mathbf{r},0\ \mathrm{T})$] from the ones measured at different magnetic fields [$g_0(\mathbf{r},B)$] in the same area of (a) and with $B$ = 0.5, 1, 2, 4, and 6 T, respectively (see text). The red circles mark the same position that has pinned a vortex under different magnetic fields. All the differential conductance mappings were recorded with $V_{bias} = 10$ mV, $I_{set} = 100$ pA. All the data were recorded at the temperature about 400 mK.
} \label{fig1}
\end{figure}

Fig.~\ref{fig1}(a) shows topographic image of the cleaved surface of FeTe$_{0.55}$Se$_{0.45}$, and Fig.~\ref{fig1}(b) shows an enlarged view of the marked square area in Fig.~\ref{fig1}(a) with atomic resolution. One can see that atoms on the surface form a well organized square lattice. The Se atoms have lower height and the Te ones have higher height. No interstitial irons can be observed on the surface, and they would behave as very bright spots locating within the plaques of four neighboring Te/Se atoms if they existed \cite{SH Pan}. In type-II superconductors, quantized vortices will be formed with applying of magnetic field stronger than the lower critical field. Due to the formation of the vortex bound states and the depairing of the Cooper pairs in the normal-state vortex core area, the density of states (DOSs) at zero energy within vortex cores should be higher than other superconducting areas without vortices where DOSs of normal electrons are gapped or partially gapped. Therefore, the vortices can be imaged by STM by zero-bias differential conductance mapping [$g_0(\mathbf{r},B)$] with $\mathbf{r}$ the location vector and $B$ the applied magnetic field. Here the measured differential conductance is proportional to the local DOSs approximately. Fig.~\ref{fig1}(d) shows the measured zero-bias differential conductance mapping $g_0(\mathbf{r},0.5\ \mathrm{T})$. Compared with $g_0(\mathbf{r},0\ \mathrm{T})$ shown in Fig.~\ref{fig1}(c), one can see that some spots are induced by the impurity bound states at zero-bias instead of the vortices, and this has also been seen in previous studies in FeTe$_{1-x}$Se$_x$ \cite{Machida}. In order to get a clear view of the vortices, we subtract $g_0(\mathbf{r},0\ \mathrm{T})$ from $g_0(\mathbf{r},B)$ to remove some impurity induced states near zero bias. The obtained difference mappings of $g_0(\mathbf{r},B)-g_0(\mathbf{r},0\ \mathrm{T})$ are shown in Fig.~\ref{fig1}(e-i) at different magnetic fields.

\begin{figure}[H]
\centering
\includegraphics[width=8cm]{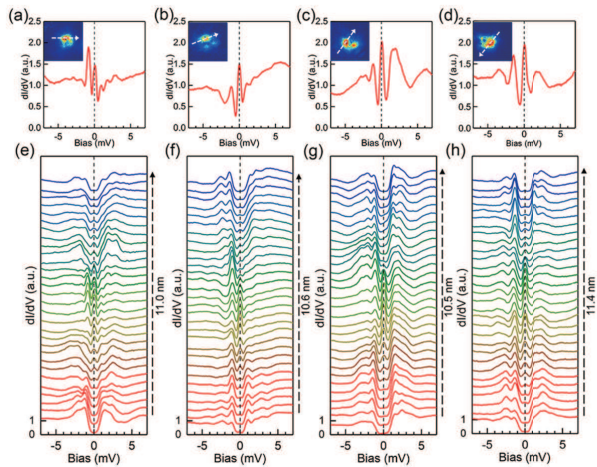}
\caption{Zero-bias conductance peaks observed in different vortices at about 400 mK. (a-d) Tunneling spectra measured in the center of the related vortex whose zero-bias differential conductance image is shown in the inset. The imaging area for each vortex has the dimensions of 20 nm $\times$ 20 nm and is recorded with $V_{bias} = 10$ mV and $I_{set} = 100$ pA. The magnetic fields for different measurements are (a) 0.5 T, (b) 1.0 T, (c,d) 2.0 T. (e-h) Tunneling spectra measured along the white dashed arrows in the insets of (a-d), respectively.
} \label{fig2}
\end{figure}

By measuring hundreds of vortices in many different areas on different samples at different magnetic fields, we can observe ZBCPs at the centers of some vortex cores. Some typical spectra as the examples are shown in Fig.~\ref{fig2}(a-d), and they are measured at the centers of the vortices whose images are shown in the insets. Fig.~\ref{fig2}(e-h) show four sets of tunnel spectra measured across the selected vortices along the white dashed arrows in the insets of Fig.~\ref{fig2}(a-d), respectively. We can see that these ZBCPs actually do not split but the intensities of these peaks shrink when the STM tip moves away from the vortex core center. The ZBCPs present in the range with a radius of about 4-5 nm around the vortex center from our experiments. The results are consistent with previous observations of ZBCPs in FeTe$_{1-x}$Se$_x$ \cite{HJ Gao,Machida}. When the spectra are measured far away from the vortex core center, they behave as a fully gapped feature with one or two pairs of coherence peaks at about 2 meV. The presence of the non-split differential conductance peaks locating exactly at zero bias can not be interpreted as the CdGM states which were observed in our previous work \cite{CdGM NC}. These CdGM state peaks generally exhibit half integer energy values with the first energy level locating at around 0.3 to 0.5 meV.

\begin{figure}[H]
\centering
\includegraphics[width=6cm]{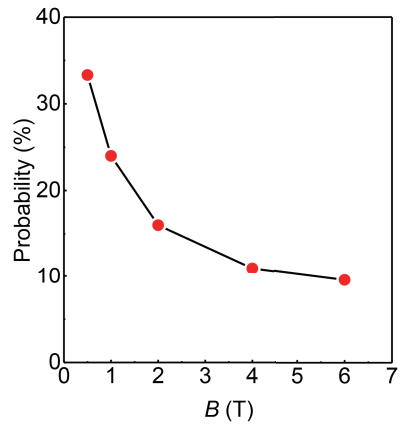}
\caption{Statistical probabilities of the vortices with ZBCPs over all the measured vortices under different magnetic fields. The total vortex numbers for statics are 15, 33, 25, 160 and 240 for the magnetic fields of 0.5, 1, 2, 4 and 6 T, respectively.
} \label{fig3}
\end{figure}

In order to investigate the influence of the magnetic field, we measured a large number of vortices under different magnetic fields in the same area shown in Fig.~\ref{fig1}(a). We do the statistic analysis on the probability of vortex cores with ZBCPs, and the result is shown in Fig.~\ref{fig3}. At the magnetic fields below 4 T, we have carefully measured the tunneling spectrum at each vortex core center. Here we have to mention that since there are many vortices that have single peaks close to zero energy, in this case, we take a criterion of the bias voltage window from $-0.1$ to $+0.1$ mV, to determine whether the vortex has a ZBCP. In other words, if the spectrum at the vortex center shows a single dominant peak close to zero bias and within this bias voltage window, we regard them all as the vortices with ZBCPs. For the situation at 0.5 and 1 T, the vortices are diluted in the field of view. We do the measurements for several times through re-adding the magnetic field in the same area in order to increase the sampling number. The vortices change their positions when we reapply the magnetic field. While at the magnetic fields of 4 and 6 T, we did not measure the spatial dependent spectra crossing each vortex core because the number of the vortices in the area becomes quite large. What we measured are the differential conductance mappings at bias voltage between $-1.0$ and $+1.0$ mV with increment of 0.2 mV. Then we convert the differential conductance mappings at different energies to the tunneling spectra and judge whether the vortex has a ZBCP. From the statistic result in Fig.~\ref{fig3}, one can see that the probability of vortices with ZBCPs decrease quickly with increase of magnetic field, which means that magnetic field may be an important factor to affect the emergence of ZBCP within the vortex core center.

\begin{figure}[H]
\centering
\includegraphics[width=8cm]{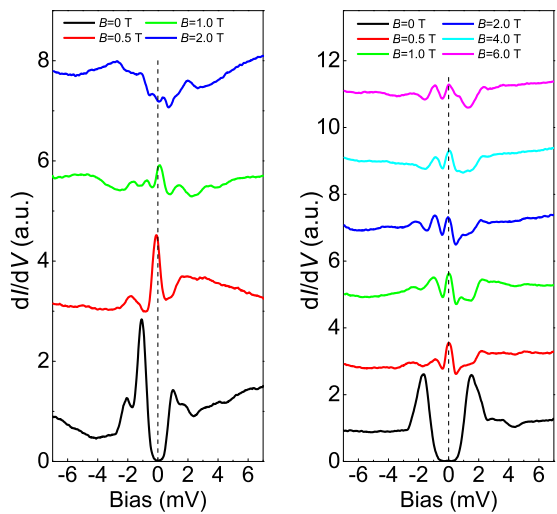}
\caption{The influence of magnetic field on the ZBCPs on the tunneling spectra taken at vortex cores. For each panel, the tunneling spectra were measured in the center of the vortices staying in the same area under different magnetic fields.
} \label{fig4}
\end{figure}

When the magnetic field varies, we can see from the vortex images in Fig.~\ref{fig1}(e-i) that some vortices could remain at the same positions by some weak pinning centers, e.g. the red circles show the same position that has pinned a vortex under different magnetic fields. These pinned vortices thus give us opportunities to study the influence of the magnetic field to the ZBCPs in the vortex cores of FeTe$_{0.55}$Se$_{0.45}$ and exclude other local effects. We find that there are some vortices that exhibit ZBCPs at low magnetic field. And by further increasing the magnetic field, although these vortices remain at the same positions, the general shape of spectrum changes with magnetic field. Figure~\ref{fig4} shows the results measured at the centers of two such vortices appearing at different locations. In Fig.~\ref{fig4}(a), the tunneling spectrum at the vortex core center shows a dominant peak at $-0.1$ meV (close to zero) at 0.5 T. At the magnetic field of 1 T, the central peak remains but moves a little to a positive energy. When the magnetic field increases to 2 T, one can see that the central peak disappears and two tiny peaks show up at around $\pm$0.35 meV, which seems to suggest the conductance peak near zero bias splits upon using a higher magnetic field. However, in another vortex [see Fig.~\ref{fig4}(b)], the zero-bias peak remains as a single one even when the magnetic field reaches 6 T. While the weight of the zero-bias peak decreases with increase of field. The evolution of the tunneling spectra at the vortex core center with the change of the magnetic field could possibly explain why the fraction of the vortices with ZBCPs becomes smaller when the magnetic field is increased. Our results show that when the magmatic field goes higher, the zero energy mode in some vortices could be destroyed and the CdGM states appear, while in some other vortices the ZBCPs still remain up to 6 T.

\begin{figure}[H]
\centering
\includegraphics[width=8cm]{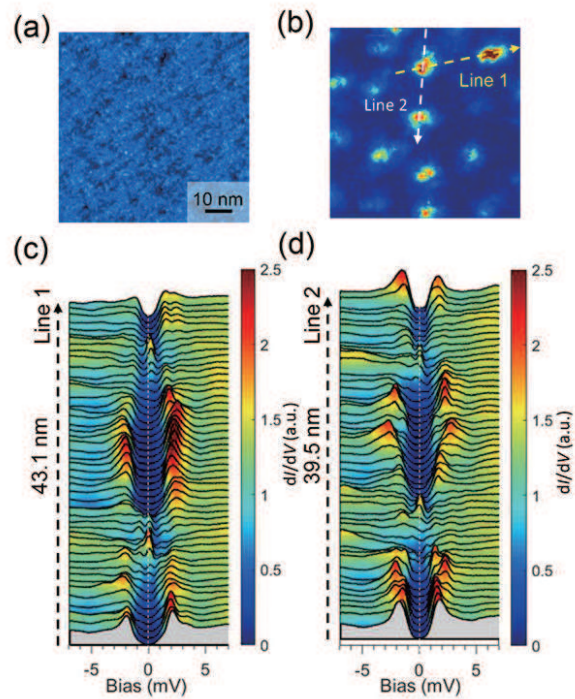}
\caption{Vortices with ZBCPs at 6 T. (a) Topographic image measured in an area of 70 nm $\times$ 70 nm. $V_{ bias} = 1$ V, $I_{set} = 10$ pA. (b) Differential conductance mapping at zero bias measured in the same area of (a) at $B = 6$ T and temperature around 400 mK. $V_{bias} = 10$ mV, $I_{set} = 100$ pA. (c,d) Tunneling spectra measured along the marked line 1 and 2 in (b).
} \label{fig5}
\end{figure}

When the magnetic field is increased, the average distance between the vortices becomes smaller. Meanwhile the repulsive interaction between vortices becomes stronger. To further evaluate the influence of magnetic field to the ZBCP, we specially take a careful look to the vortices at 6 T at which field the vortices should have stronger interactions. The fraction for observing the vortices with ZBCPs is only about 10\% at 6 T, so these vortices are randomly distributed and do not usually stay close to each other. However, we do find in one case that there are three neighboring vortices which all exhibit ZBCPs. Here we show the topographic image of the measured area in Fig.~\ref{fig5}(a), and  Fig.~\ref{fig5}(b) shows the $dI/dV$ mapping of vortices at 6 T measured in this area. The average distance is about 20 nm between the nearest neighboring vortices. We measured spatially dependent spectra by following the two arrowed lines in Fig.~\ref{fig5}(b), and the results are shown in Fig.~\ref{fig5}(c,d). One can see that these three neighbored vortices host the ZBCPs, and these peaks almost do not shift when the tip moves away from the vortex core center. The results indicate that even though the nearest neighboring vortices have strong interactions, the ZBCPs can still exist in these vortex centers.

\begin{figure}[H]
\centering
\includegraphics[width=7cm]{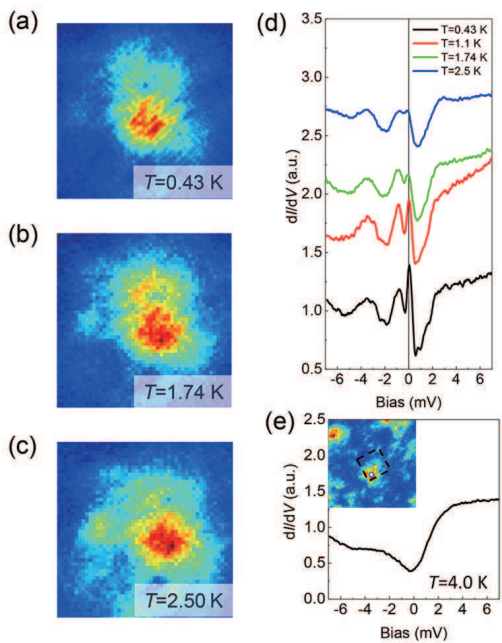}
\caption{Temperature dependence of the spectra with ZBCP at low temperature. (a-c) $dI/dV$ mappings measured in the same area (14 nm $\times$ 14 nm) at different temperatures under a magnetic field of 2 T. A single vortex locates almost at the same positions in (a-c). (d) Tunneling spectra measured at the vortex core centers at different temperatures. (e) Tunneling spectrum measured at another vortex core center at $T$ = 4.0 K. The insert shows the $dI/dV$ mapping of the measured area (50 nm $\times$ 50 nm). The black dashed square frame represents the same area in (a-c), while the tiny blue circle marks the position where the tunneling spectrum is taken at $T$ = 4.0 K. One can see that the location of the vortex changes a little from the initial positions at 0.43, 1.74, 2.50 K (a-c) from that at 4.0 K.
} \label{fig6}
\end{figure}

To further investigate the characteristics of the ZBCPs, we measured the temperature dependence of the tunneling spectra at the center of a vortex core, and the results are shown in Fig.~\ref{fig6}. In Fig.~\ref{fig6}(d), one can see the differential conductance peak locating almost at zero bias on the spectrum taken in the center of a vortex shown in Fig.~\ref{fig6}(a) at 0.43 K and 2 T. With increase of temperature to 2.5 K, we try to measure the tunneling spectrum in the center of the vortex which stays at almost the same position as that of the vortex at 0.43 K. The intensity of the ZBCP is lowered down obviously from that shown in Fig.~\ref{fig6}(d), and the side peak at negative bias energy is also weakened simultaneously. The ZBCP is very weak on the tunneling spectra taken at 2.5 K. When the temperature increases to 4.0 K, the vortex shifts a little bit. However fortunately from the initial place, we can observe a vortex nearby as shown in the inset of Fig.~\ref{fig6}(e), and the tunneling spectrum measured in the center of this vortex is shown in Fig.~\ref{fig6}(e). One can see that the feature of the ZBCP and the side peak are absent on the spectrum measured in the center of the vortex at 4.0 K. Considering the strong suppression of ZBCP at 2.5 K, we argue that the ZBCP may be completely suppressed at about 4.0 K as evidenced by the spectrum measured in another vortex nearby, as shown in Fig.~\ref{fig6}(e).

\begin{figure}[H]
\centering
\includegraphics[width=7cm]{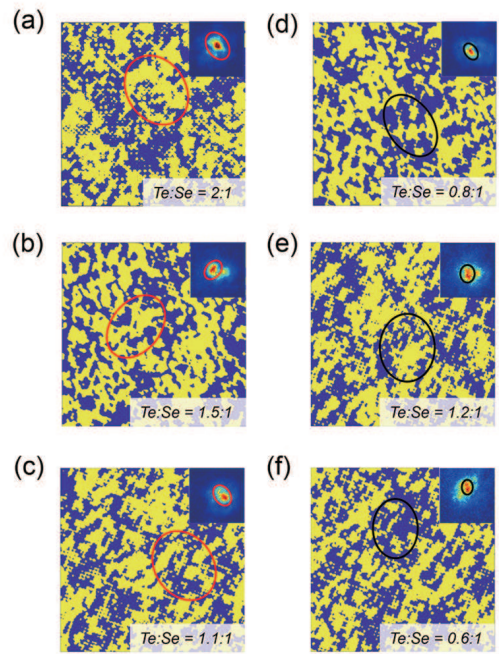}
\caption{Local Te/Se concentrations near the vortices with or without ZBCPs. The dark blue color covers the area of Se atoms on the surface, while the bright yellow color covers the area of Te atoms. The red ellipses show roughly the major areas of the observed vortex cores whose images are shown in the insets by differential conductance mapping at zero bias. (a-c) The Te/Se distributions near vortices with ZBCPs. (d-f) The Te/Se distributions near vortices without ZBCPs.
} \label{fig7}
\end{figure}

In order to investigate the influence of the local chemical potential to the presence of the ZBCP at vortex center, we measured the atomical-resolved topography in the region near the vortex core. The Te or Se atoms can be distinguished by the height of the atoms. The proportion of Te:Se in a large area is close to the nominal composition 0.55:0.45, but such proportion changes slightly in different local areas. In the main panels of Fig.~\ref{fig7}, we show the topography measured near the vortices by binary color images, i.e., the bright yellow (dark blue) color to fill the area of Te (Se) atoms on the surface with higher (lower) heights. Figures~\ref{fig7}(a-c) show the topographic images measured near the vortices with ZBCPs, while Fig.~\ref{fig7}(d-f) show the ones measured near the vortices without ZBCPs but with CdGM states. One can see that the result doesn't show any direct relationship of Te/Se concentrations to the emergence of ZBCPs in the vortex cores.

\section{Discussions}

As presented above, we have measured the tunneling spectra in a large number of vortices in FeTe$_{0.55}$Se$_{0.45}$, the ZBCPs are only observed in some vortices. The feature of these ZBCPs is different from that of the CdGM states observed in the vortex cores in superconductors with a large value of $E_F$ \cite{Hess 1,Hess 2}, which is consistent with previous reports \cite{HJ Gao,Machida}. In our previous work \cite{CdGM NC}, we observed the discrete CdGM states due to very small $E_F$ in some vortices. These CdGM states can also be observed in the vortex cores with or without ZBCPs. Besides, the detailed structures and backgrounds of the spectrum in the center of vortex are very complex at energies near the Fermi energy. In another set of measurements \cite{Machida} in some vortices on the same material taken at a lower effective temperature of about 85 mK, the authors find very sharp ZBCPs accompanied by some satellite peaks which can not be explained by a straightforward picture. In our measurements, the temperature is about 0.4 K, the thermal broadening effect on these peaks may mix the ZBCP with satellite peaks and may slightly change the exact maxima energy of the ZBCP. In this case, it maybe reasonable for us to regard the in-gap bound state peaks at energies within the window $\pm0.1$ meV as the ZBCPs. With this point of view, the single peak observed in some vortex cores near zero energy in our previous work \cite{CdGM NC} may also be the ZBCP mixed with other bound-state peaks. In addition, the previous measurements were carried out at the magnetic field of 4 T \cite{CdGM NC}, and the fraction for observing the vortices with ZBCPs is only about 10\% according to the statistic result in Fig.~\ref{fig3}. Hence, it is understandable that none of the measured vortices presented in our previous work exhibit the differential conductance peaks exactly at zero bias.

For all the results measured on the vortices of FeTe$_{0.55}$Se$_{0.45}$, we found that only a fraction of them exhibit ZBCPs. Since the topological surface states have been predicted and proved in some way in this material \cite{ARPES2}, it is really strange why the ZBCPs can only be observed in some of the vortices if they are regarded as the Majorana bound state at zero energy \cite{HJ Gao}. The ZBCPs can be observed in the vortex center at the magnetic field up to 6 T, however, the ratio of the vortices with ZBCPs decrease rapidly with increase of magnetic field, which is consistent with a previous work \cite{Machida}. Our present experiments reveal that the single ZBCP is strongly suppressed or even splits into two peaks at high magnetic field. This may explain why the fraction for observing the vortices with ZBCPs at a high magnetic field is low. And in this point of view, some of the ZBCPs, which split at high magnetic field, may not be the Majorana zero-energy modes. Even for the Majorana zero-energy modes, a theoretical proposal is that the strong interaction between the vortices at high magnetic field could lead to the suppression to these fragile modes \cite{interact 1,interact 2}. Interestingly, we found three neighboring vortices that all exhibit ZBCPs even at a magnetic field of 6 T. This may suggest that the disordered arrangement of the Majorana distribution and the Majorana hybridization lead to the decreasing fraction of ZBCPs\cite{Chiu}. For the temperature evolution, the ZBCP becomes very weak at 2.5 K, and may disappear completely at about 4 K. Therefore these ZBCPs are very fragile to the thermal effect. Moreover, we did not find the direct relationship between the local Te/Se concentration on the top surface and the occurrence of the ZBCP within the vortex cores.

\section{Conclusion}

In conclusion, through measuring the scanning tunneling spectra within the vortex cores in FeTe$_{0.55}$Se$_{0.45}$, we have confirmed the discovery of ZBCPs in some vortices. The ZBCP does not split when moving away from the vortex core center, which is different from that expected by the CdGM states in the vortex cores in most superconductors with large $E_F$, thus it is most likely the Majorana mode. The temperature dependence of the spectra measured in the vortex core center shows that the ZBCP may be suppressed completely at about 4 K, indicating a thermal-fragile feature of the ZBCP. Furthermore, we find that a large magnetic field can suppress the fraction of the vortices with ZBCPs, but the ZBCP can sometime exist at a magnetic field as high as 6.0 T. We did not find any close relationship between the local Te/Se concentration and the presence of the ZBCP within vortex cores. Our results will help to elaborate the origin of these ZBCPs found in the vortex core states in FeTe$_{0.55}$Se$_{0.45}$.

\begin{acknowledgments}
We acknowledge useful discussions with Tetsuo Hanaguri, Ching-Kai Chiu, Tadashi Machida, and Hong Ding. This work was supported by National Key R\&D Program of China (grant number: 2016YFA0300401), National Natural Science Foundation of China (grant number: 11534005), and the Strategic Priority Research Program of Chinese Academy of Sciences (grant number: XDB25000000).
\end{acknowledgments}

$^*$ huanyang@nju.edu.cn

$^\dag$ hhwen@nju.edu.cn


\begin{thebibliography}{40}

\bibitem{CdGM} C. Caroli, P. G. De Gennes, and J. Matricon, \textit{Bound Fermion states on a vortex line in a type II superconductor}, Phys. Lett. \textbf{9}, 307 (1964).

\bibitem{vortex theory 2} N. Hayashi, T. Isoshima, M. Ichioka, and K. Machida, \textit{Low-lying quasiparticle excitations around a vortex core in quantum limit}, Phys. Rev. Lett. \textbf{80}, 2921 (1998).

\bibitem{Hess 1} H. F. Hess, R. B. Robinson, R. C. Dynes, J. M. Valles, and J. V. Waszczak, \textit{Scanning-Tunneling-Microscope observation of the Abrikosov flux lattice and the density of states near and inside a fluxoid}, Phys. Rev. Lett. \textbf{62}, 214 (1989).

\bibitem{Hess 2} H. F. Hess, R. B. Robinson, and J. V. Waszczak, \textit{Vortex-core structure observed with a scanning tunneling microscope}, Phys. Rev. Lett. \textbf{64}, 2711 (1990).

\bibitem{vortex theory 1} F. Gygi and M. Schl\"{u}ter, \textit{Self-consistent electronic structure of a vortex line in a type-II superconductor}, Phys. Rev. B \textbf{43}, 7609 (1991).

\bibitem{FTS Tc} R. Khasanov, M. Bendele, A. Amato, P. Babkevich, A. T. Boothroyd, A. Cervellino, K. Conder, S. N. Gvasaliya, H. Keller, H. H. Klauss, H. Luetkens, V. Pomjakushin, E. Pomjakushina, and B. Roessli, \textit{Coexistence of incommensurate magnetism and superconductivity in ${\text{Fe}}_{1+y}{\text{Se}}_{x}{\text{Te}}_{1\ensuremath{-}x}$}, Phys. Rev. B \textbf{80}, 140511 (2009).

\bibitem{ARPES FTS1} Y. Lubashevsky, E. Lahoud, K. Chashka, D. Podolsky, and A. Kanigel, \textit{Shallow pockets and very strong coupling superconductivity in FeSe$_{x}$Te$_{1-x}$}, Nat. Phys. \textbf{8}, 309 (2012).

\bibitem{ARPES FTS2} S. Rinott, K. B. Chashka, A. Ribak, E. D. L. Rienks, A. Taleb-Ibrahimi, P. Le Fevre, F. Bertran, M. Randeria, and A. Kanigel, \textit{Tuning across the BCS-BEC crossover in the multiband superconductor Fe$_{1+y}$Se$_{x}$Te$_{1-x}$: An angle-resolved photoemission study}, Sci. Adv. \textbf{3}, e1602372 (2017).

\bibitem{CdGM NC} M. Chen, X. Chen, H. Yang, Z. Du, X. Zhu, E. Wang, and H.-H. Wen, \textit{Discrete energy levels of Caroli-de Gennes-Matricon states in quantum limit in FeTe$_{0.55}$Se$_{0.45}$}, Nat. Commun. \textbf{9}, 970 (2018).

\bibitem{ARPES1} Z. Wang, P. Zhang, G. Xu, L. K. Zeng, H. Miao, X. Xu, T. Qian, H. Weng, P. Richard, A. V. Fedorov, H. Ding, X. Dai, and Z. Fang, \textit{Topological nature of the FeSe$_{0.5}$Te$_{0.5}$ superconductor}, Phys. Rev. B \textbf{92}, 115119 (2015).

\bibitem{FST Band} G. Xu, B. Lian, P. Tang, X.-L. Qi, and S.-C. Zhang, \textit{Topological superconductivity on the surface of Fe-based superconductors}, Phys. Rev. Lett. \textbf{117}, 047001 (2016).

\bibitem{HuJP} X. Wu, S. Qin, Y. Liang, H. Fan, and J. Hu, \textit{Topological characters in Fe(Te$_{1-x}$Se$_{x}$) thin films}, Phys. Rev. B \textbf{93}, 115129 (2016).

\bibitem{ARPES2} P. Zhang, K. Yaji, T. Hashimoto, Y. Ota, T. Kondo, K. Okazaki, Z. Wang, J. Wen, G. D. Gu, H. Ding, and S. Shin, \textit{Observation of topological superconductivity on the surface of an iron-based superconductor}, Science \textbf{360}, 182 (2018).

\bibitem{SH Pan} J. X. Yin, Z. Wu, J. H. Wang, Z. Y. Ye, J. Gong, X. Y. Hou, L. Shan, A. Li, X. J. Liang, X. X. Wu, J. Li, C. S. Ting, Z. Q. Wang, J. P. Hu, P. H. Hor, H. Ding, and S. H. Pan, \textit{Observation of a robust zero-energy bound state in iron-based superconductor Fe(Te,Se)}, Nat. Phys. \textbf{11}, 543 (2015).

\bibitem{ZQ Wang} K. Jiang, X. Dai, and Z. Wang, \textit{Quantum anomalous vortex and Majorana zero mode in iron-based superconductor Fe(Te,Se)}, Phys. Rev. X \textbf{9}, 011033 (2019).

\bibitem{Read and Green} N. Read and D. Green, \textit{Paired states of fermions in two dimensions with breaking of parity and time-reversal symmetries and the fractional quantum Hall effect}, Phys. Rev. B \textbf{61}, 10267 (2000).

\bibitem{Fu and Kane} L. Fu and C. L. Kane, \textit{Superconducting proximity effect and Majorana fermions at the surface of a topological insulator}, Phys. Rev. Lett. \textbf{100}, 096407 (2008).

\bibitem{HJ Gao} D. Wang, L. Kong, P. Fan, H. Chen, S. Zhu, W. Liu, L. Cao, Y. Sun, S. Du, J. Schneeloch, R. Zhong, G. Gu, L. Fu, H. Ding, and H.-J. Gao, \textit{Evidence for Majorana bound states in an iron-based superconductor}, Science \textbf{362}, 333 (2018).

\bibitem{half  intefer level shift} L. Kong, S. Zhu, M. Papaj, L. Cao, H. Isobe, W. Liu, D.Wang, P. Fan, H. Chen, Y. Sun, S. Du, J. Schneeloch, R. Zhong, G. Gu, L. Fu, H.-J. Gao, and H. Ding, \textit{Observation of half-integer level shift of vortex bound states in an iron-based superconductor}, arXiv:1901.02293.

\bibitem{2e2/h FTS} S. Zhu, L. Kong, L. Cao, H. Chen, S. Du, Y. Xing, W. Liu, D. Wang, C. Shen, F. Yang, J. Schneeloch, R. Zhong, G. Gu, L. Fu, Y.-Y. Zhang, H. Ding, H.-J. Gao, \textit{Observation of Majorana conductance plateau by scanning tunneling spectroscopy}, arXiv:1904.06124.

\bibitem{quantized plateau 1} K. T. Law, P. A. Lee, and T. K. Ng, \textit{Majorana Fermion Induced Resonant Andreev Reflection}, Phys. Rev. Lett. \textbf{103}, 237001 (2009).

\bibitem{quantized plateau 2} K. Flensberg, \textit{Tunneling characteristics of a chain of Majorana bound states}, Phys. Rev. B \textbf{82}, 180516 (2010).

\bibitem{quantized plateau 3} M. Wimmer, A. R. Akhmerov, J. P. Dahlhaus, and C. W. J. Beenakker, \textit{Quantum point contact as a probe of a topological superconductor}, New J. Phys. \textbf{13}, 053016 (2011).

\bibitem{P Zhang NP} P. Zhang, Z. Wang, X. Wu, K. Yaji, Y. Ishida, Y. Kohama, G. Dai, Y. Sun, C. Bareille, K. Kuroda, T. Kondo, K. Okazaki, K. Kindo, X. Wang, C. Jin, J. Hu, R. Thomale, K. Sumida, S. Wu, K. Miyamoto, T. Okuda, H. Ding, G. D. Gu, T. Tamegai, T. Kawakami, M. Sato, and S. Shin, \textit{Multiple topological states in iron-based superconductors}, Nat. Phys. \textbf{15}, 41 (2019).

\bibitem{LiFeOHFeSe 1} Q. Liu, C. Chen, T. Zhang, R. Peng, Y.-J. Yan, C.-H.-P. Wen, X. Lou, Y.-L. Huang, J.-P. Tian, X.-L. Dong, G.-W. Wang, W.-C. Bao, Q.-H. Wang, Z.-P. Yin, Z.-X. Zhao, and D.-L. Feng, \textit{Robust and clean Majorana zero mode in the vortex core of high-temperature superconductor (Li$_{0.84}$Fe$_{0.16}$)OHFeSe}, Phys. Rev. X \textbf{8}, 041056 (2018).

\bibitem{LiFeOHFeSe 2} C. Chen, Q. Liu, T. Z. Zhang, D. Li, P. P. Shen, X. L. Dong, Z.-X. Zhao, T. Zhang, D. L. Feng, \textit{Quantized conductance of Majorana zero mode in the vortex of the topological superconductor (Li$_{0.84}$Fe$_{0.16}$)OHFeSe}, arXiv:1904.04623.

\bibitem{Growth} Y. Liu, and C. T. Lin, \textit{A Comparative study of Fe$_{1+\delta}$Te$_{1-x}$Se$_{x}$ single crystals grown by Bridgman and self-flux techniques}, J. Supercond. Nov. Magn. \textbf{24}, 183 (2011).

\bibitem{Machida} T. Machida, Y. Sun, S. Pyon, S. Takeda, Y. Kohsaka, T. Hanaguri, T. Sasagawa, and T. Tamegai, \textit{Zero-energy vortex bound state in the superconducting topological surface state of Fe(Se,Te)}, Nat. Mater. \textbf{18}, 811 (2019).

\bibitem{interact 1} C.-K. Chiu, D. I. Pikulin, and M. Franz, \textit{Strongly interacting Majorana fermions}, Phys. Rev. B \textbf{91}, 165402 (2015).

\bibitem{interact 2} T. Liu and M. Franz, \textit{Electronic structure of topological superconductors in the presence of a vortex lattice}, Phys. Rev. B \textbf{92}, 134519 (2015).

\bibitem{Chiu} C.-K. Chiu, T. Machida, Y. Huang, T. Hanaguri, and F.-C. Zhang, \textit{Scalable Majorana vortex modes in iron-based superconductors}, arXiv:1904.13374.


\end{thebibliography}
\end{document}